\AtBeginDocument{}
\documentclass[aps,prl,twocolumn,superscriptaddress,calc,epsfig,10pt,floatfix, groupedaddress]{revtex4-2}
\usepackage{graphicx}
\usepackage{graphics}
\usepackage{amssymb,amsmath}
\usepackage{wasysym}
\usepackage{float}
\usepackage{physics}
\usepackage[dvipsnames]{xcolor}
\usepackage[normalem]{ulem} 
\usepackage[pdftex,breaklinks,colorlinks=true,linkcolor=blue,citecolor=blue,urlcolor=blue]{hyperref}
\usepackage{tikz}
\usetikzlibrary{calc,shapes}
\usepackage{lineno}  % Makes linenumbers work well. Have to manually adjust \internallinenumbers right above caption call for figures

\usepackage{siunitx}

\newcommand{\mitCUAaddress}{Department of Physics, MIT-Harvard Center for Ultracold Atoms, and Research Laboratory of Electronics, MIT, Cambridge, Massachusetts 02139, USA}

\begin{document}
\thickmuskip = 3mu

\title{Revealing Hidden Correlations in a Fermi-Hubbard system via Interaction Ramps}

\author{Botond Oreg}
\thanks{Present address: Atom Computing, Inc., 2500 55th St, Boulder, CO 80301}
\author{Carter Turnbaugh}
\author{Jens Hertkorn}
\author{Ningyuan Jia}
\thanks{Present address: QuEra Computing, Inc., 1380 Soldiers Field Road,
Boston, MA 02135}
\author{Martin Zwierlein}
\date{\today}
\affiliation{\mitCUAaddress}

\begin{abstract}
We observe an enhanced visibility of charge-density-wave correlations in a cold-atom realization of the attractive Hubbard model following a rapid boost of the interaction strength.
The interaction boost associates nonlocal pairs into doublons which mark the center of mass of the original pairs.
The enhancement is largest in the strongly correlated regime where pairing is nonlocal.
We distinguish the unpaired Fermi liquid from the pseudogap phase of preformed pairs by analyzing atom-resolved spin-charge correlations after the ramp.
The technique we establish here may facilitate the observation of exotic forms of pair order in spin-imbalanced systems, and of stripe order in the dual case of the doped repulsive Hubbard model.
\end{abstract}

\maketitle

%%%%%%%%%%%%%%%%%%%%%%%%%%%%%%%%%%%%%%%%%%%%%%
%%%%%%%%%%%%%%% Introduction %%%%%%%%%%%%%%%%%
%%%%%%%%%%%%%%%%%%%%%%%%%%%%%%%%%%%%%%%%%%%%%%
%\section{Introduction}
\begin{figure}[t]
    \centering
    \includegraphics[width=\columnwidth]{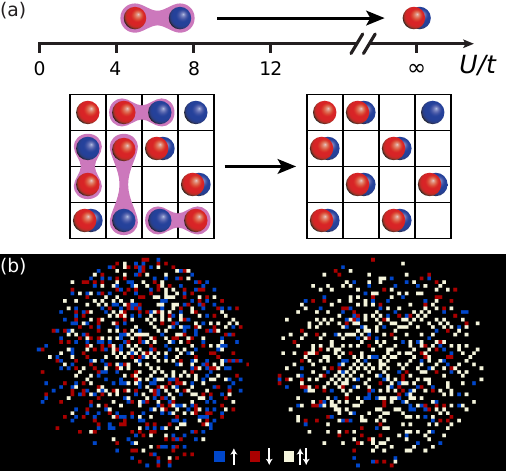}
    % \internallinenumbers
    \caption{\textbf{Rapid ramp in an attractive Fermi-Hubbard gas.}
    \textbf{(a)}~The attraction strength between fermions can be dynamically tuned near a Feshbach resonance. 
    A rapid ramp of the interaction strength associates nonlocal pairs into local doublons.
    This reveals otherwise hidden order between pairs, e.g. a charge-density-wave pattern.
    \textbf{(b)}~Two experimental realizations of the same, strongly correlated, Fermi-Hubbard system ($U/t = 5.8(3)$) without (left) and with (right) the rapid ramp applied.
    }
    \label{fig:Introduction}
\end{figure}

The interplay of fermion pairing and charge order is a central theme in the study of strongly correlated electron systems. High-temperature cuprates~\cite{chang2012direct} and bismuthates~\cite{harris2020charge}, as well as transition-metal dichalcogenides~\cite{morosan2006superconductivity}, all exhibit superconductivity in close proximity to charge-density-wave order, raising fundamental questions about the relationship between pairing and spatial symmetry breaking~\cite{fradkin2015colloquium}. The attractive Hubbard model provides a paradigm for the study of such intertwined order: at half-filling, superfluid and charge-density-wave states are degenerate. The model can be realized with ultracold atoms in optical lattices, and fermion pairing as well as strong charge-density correlations have been observed with single-atom-resolving microscopy~\cite{mitra2018quantum, brown2020angle, Hartke2023Direct}.
However, in the strongly interacting regime where the interaction strength is on the order of the bandwidth and correlations should be largest, the pair size is comparable to the interparticle spacing~\cite{Randeria1992Pairing, Trivedi1995Deviations, Hartke2023Direct}. This nonlocal character hinders the direct detection both of the pairs themselves and of their spatial order.

The dynamical tunability of parameters in a cold-atom realization offers a powerful route to access such otherwise hidden order. In strongly interacting bulk systems, rapid interaction sweeps before imaging have revealed pair condensation~\cite{Regal2004Observation, Zwierlein2004Condensation} and the formation of vortex lattices in superfluids~\cite{Zwierlein2005Vortices} and supersolids~\cite{Casotti2024Observation}. In lattice systems, dynamical control combined with local readout has enabled the detection of otherwise hidden observables such as current and kinetic energy~\cite{impertro2024local}, and proposals exist to detect general observables~\cite{tran2023measuring} and even $d$-wave pairing~\cite{schlomer2024local, mark2025efficiently}. 
In the repulsive Hubbard model, which stands in direct correspondence to the attractive one~\cite{Shiba1972, Emery1976, Moreo2007, ho2009quantum}, dynamical ramps of the lattice depth have been utilized to suppress doublon-hole fluctuations~\cite{Kale2022Schrieffer}.

In this work, we introduce an interaction quenching technique to probe the spatial order of fermion pairs in the attractive Hubbard gas, as illustrated in Fig.~\ref{fig:Introduction}(a).
Starting with an equilibrium state, we rapidly increase the attraction strength via a Feshbach resonance immediately prior to imaging.
This quench projects extended, nonlocal pairs onto local doublons.
We observe that the ramp significantly enhances the visibility of charge-density-wave (CDW) order in the regime where pairs are initially nonlocal.
Following the ramp, the doublon density serves as a local proxy for the presence of pairs, allowing us to characterize the crossover from a Fermi liquid to the paired regime.

We realize the two-dimensional attractive Hubbard model with a spin-balanced Fermi gas of \textsuperscript{40}K atoms (density per site $n \sim 0.8$, temperature $T/t \sim 0.5$) in a $4.3(2) E_r$ deep sinusoidal square lattice, where $E_r$ is the recoil energy.
In the lowest band, the atoms are described by the Hubbard Hamiltonian
\begin{equation}
    H = -t \sum\limits_{\langle i j \rangle, \sigma} c^\dagger_{i, \sigma} c_{j, \sigma} - U \sum\limits_{i} n_{i, \uparrow} n_{i, \downarrow}
\end{equation}
with tunneling $t=h \times \SI{340}{Hz}$.
The attraction $U$ is controlled via the applied magnetic field near a Feshbach resonance at $\SI{202.1}{G}$.

The interaction quench consists of a linear magnetic field ramp from the initial value to the resonance in $\SI{1}{ms}$ ($\approx 0.3 \, h/t$), as shown in Fig.~\ref{fig:Introduction}(a)~\footnote{
For our trap geometry, the post-ramp interaction energy is $U = h \times \SI{12.7}{kHz} = 37\,t$~\cite{busch1998two, idziaszek2006analytical, hartke2022quantum}.
The probability of a pair to be nonlocal scales with $(t/U)^2 \sim 7\times10^{-4}$, which is negligible.}.
Immediately after, the lattice depth is increased to ${\sim}\,70\,E_r$ in $\SI{0.1}{ms}$ to freeze the atom distribution, and
the cloud is imaged with single-site and spin resolution~\cite{Hartke2020Doublon, Hartke2023Direct}.
Figure~\ref{fig:Introduction}(b) compares typical snapshots without (left) and with (right) the interaction ramp: the doublon density and checkerboard ordering characteristic of CDW correlations are visibly enhanced.

%%%%%%%%%%%%%%%%%%%%%%%%%%%%%%%%%%%%%%%%%%%%%%%%%
%%%%%%%%%%%%%%% Charge Behavior %%%%%%%%%%%%%%%%%
%%%%%%%%%%%%%%%%%%%%%%%%%%%%%%%%%%%%%%%%%%%%%%%%%
%\section{Charge Behavior}
\begin{figure}[t]
	\centering
	\includegraphics[width=\columnwidth]{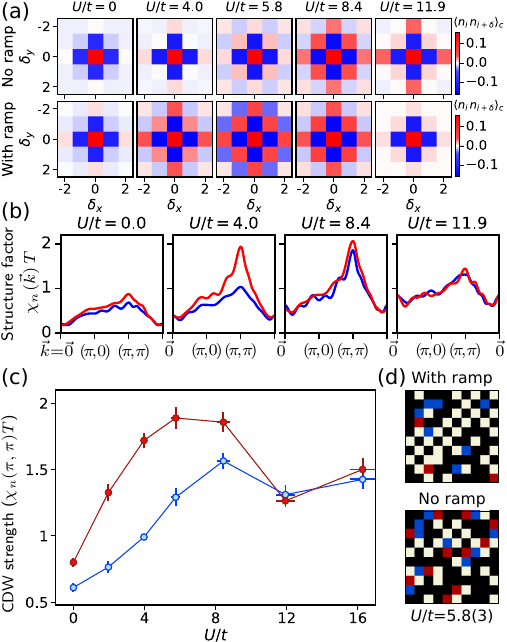}
	% \internallinenumbers
	\caption{
        \textbf{Revealing charge-density-wave (CDW) correlations with the rapid ramp.}
        \textbf{(a)}~Density-density correlation maps as a function of interaction strength $U/t$ without (upper) and with (lower) the rapid ramp.
        % The range and magnitude of the charge-density-wave (CDW) correlations increases with the rapid ramp as the pairs are associated for large enough $U/t$ values.
        \textbf{(b)}~Fourier transforms of the density correlation maps at various interaction strengths along the trace $(0,0) \rightarrow (\pi,0) \rightarrow (\pi,\pi) \rightarrow (0,0)$. A peak at $(\pi, \pi)$ indicates CDW structure.
        \textbf{(c)}~The total rectified density correlations $\sum_{\delta} (-1)^{\delta_x+\delta_y} \expval{n_i n_{i+\delta}}_c$ as a function of interaction strength with (red) and without (blue) the rapid ramp.
        The CDW signal is strongly enhanced by the ramp, with the largest effect at $4 \lesssim U/t \lesssim 8$ where the gas is nearly fully paired but pairs retain a strong nonlocal character.
        \textbf{(d)}~Portions of atomic clouds for two experimental realizations at $U/t = 5.8(3)$ without (lower) and with (upper) the rapid ramp.
	}
	\label{fig:Order}
\end{figure}

To understand the origin of CDW order and the role of the ramp, we recall the structure of pairs in the attractive Hubbard model~\cite{ho2009quantum, mitra2018quantum, Hartke2023Direct}. At $t=0$ all fermions are paired in local doublons. Nonzero tunneling causes each pair to acquire a nonlocal component in which the two spins occupy neighboring sites, lowering the energy by the pair-hopping scale $J = 4 t^2/U$~\cite{Auerbach2012Interacting}. 
This delocalization is blocked when the neighboring site is already occupied, generating an effective inter-pair repulsion of order $J$ that drives CDW correlations~\cite{ho2009quantum, mitra2018quantum, Hartke2023Direct}.

The CDW order is expected to be strongest in the crossover regime where $J$ approaches $t$ and the pair size is comparable to the interparticle spacing, yet the same nonlocal character that maximizes these correlations also obscures them from detection: a pair delocalized across two sites does not contribute to the alternating doublon-hole CDW pattern. It is in this regime that we can expect the interaction quench to be most effective: by converting extended pairs into local doublons, the ramp reveals the full extent of the underlying charge order.

Figure~\ref{fig:Order} quantifies this enhancement across the full range of interaction strengths.
The density-density correlator $\expval{n_{i}n_{i+\delta}}_c$ (Fig.~\ref{fig:Order}(a)) and its Fourier transform, the density structure factor $\chi_n(\vec{k})$ (Fig.~\ref{fig:Order}(b)), both show a dramatic increase of CDW correlations for $4 \lesssim U/t \lesssim 8$.

The CDW strength, measured by $\chi_n(\pi,\pi) = \sum_{\delta}(-1)^{\delta_x+\delta_y}\expval{n_{i}n_{i+\delta}}_c$, is shown in Fig.~\ref{fig:Order}(c) with and without the ramp.
The strongest CDW order found after the ramp, at $U/t \approx 6$, significantly exceeds the maximum attainable at any $U/t$ without the ramp. This is direct evidence that the order was present before the ramp but hidden by pair delocalization.

The three interaction regimes are clearly distinguished. For strong attraction ($U/t \gtrsim 12$), pairs are already predominantly local and the ramp has no appreciable effect. In the crossover regime ($4 \lesssim U/t \lesssim 8$), the ramp produces the largest enhancement. The peak of the post-ramp CDW signal appears at smaller $U/t$ than the maximum without the ramp and coincides with the region where the highest superfluid critical temperatures are predicted~\cite{Scalettar1989Phase, Moreo1991Two, Shen1996Pseudospin, dupuis2004berezinskii, Paiva2010Fermions} and, correspondingly, the strongest precursor CDW and superfluid correlations.
For weak attraction ($U/t \lesssim 4$), the gas is a Fermi liquid whose density correlations are governed by Pauli repulsion and show no CDW structure with or without the ramp.

%%%%%%%%%%%%%%%%%%%%%%%%%%%%%%%%%%%%%%%%%%%
%%%%%%%%%%%%%%% Timescale %%%%%%%%%%%%%%%%%
%%%%%%%%%%%%%%%%%%%%%%%%%%%%%%%%%%%%%%%%%%%
\begin{figure}[hbtp]
	\centering
	\includegraphics[width=\columnwidth]{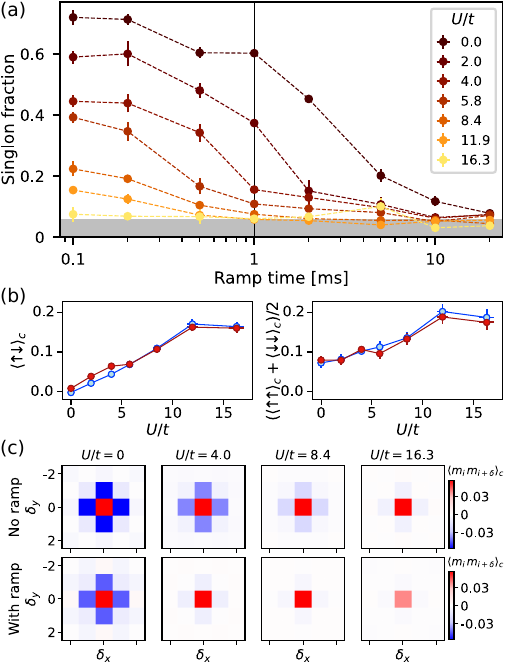}
	% \internallinenumbers
	\caption{\textbf{Calibrating and validating the ramp protocol.}
        \textbf{(a)}~Singlon fraction after ramps of various durations from a range of initial interaction strengths.
        At the $\SI{1}{ms}$ ramp time used in the experiment, the singlon fraction of a noninteracting gas is unchanged while that of a strongly interacting gas reaches the detection limit.
        \textbf{(b)}~Total unlike- ($\sum_{\delta} \expval{n_{i,\uparrow} n_{i + \delta,\downarrow}}_c$, left) and same-spin ($\sum_{\delta} \left( \expval{n_{i,\uparrow} n_{i + \delta,\uparrow}}_c + \expval{n_{i,\downarrow} n_{i + \delta,\downarrow}}_c \right)/2$, right) correlations before (blue) and after (red) a $\SI{1}{ms}$ ramp.
        \textbf{(c)}~Magnetization correlations $\expval{m_i m_{i+\delta}}_c$ ($m = n_\uparrow - n_\downarrow$) without (top) and with (bottom) the $\SI{1}{ms}$ ramp at various distances.
        In the pair-dominated regime ($U/t \gtrsim 4$), the ramp removes nonlocal magnetization correlations from extended pairs as it turns them into local doublons.
    }
	\label{fig:Timescale}
\end{figure}

The optimal timescale for the interaction ramp is set by balancing competing requirements.
The ramp must be slow enough for pair constituents to merge into doublons, yet fast enough to preserve the global pair arrangement and prevent the creation of new pairs.
The minimum time required to contract a nonlocal pair is set by the tunneling time $h/t$, while many-body ordering is governed by the pair-hopping time $h/J$~\cite{Auerbach2012Interacting}.
In the limit of large attraction ($U \gg t$), these timescales are well separated ($J \ll t$), allowing a ramp time $\tau$ that lies in between ($h/t \lesssim \tau \ll h/J$).
In the crossover regime ($J\approx t$), they become comparable, necessitating an experimental calibration of the optimal ramp time.

Fig.~\ref{fig:Timescale}(a) shows the measured singlon fraction as a function of $U/t$ and ramp time $\tau$.
For fast sweeps ($\tau \ll \SI{1}{ms}$), pairs have insufficient time to contract and the singlon fraction is unchanged.
For slow sweeps ($\tau \gg \SI{1}{ms}$), the cloud evolves adiabatically, pairing nearly all atoms and suppressing the singlon fraction to the detection limit regardless of the initial $U/t$.
We identify $\tau = \text{1 ms}$ as the optimal intermediate ramp time used in this work; here, the singlon fraction of an initially noninteracting gas remains near the fast-sweep limit, while that of a strongly attractive gas ($U/t \gtrsim 8$) drops to the background level, confirming pair localization.

We further verify that the macroscopic density distribution remains frozen by monitoring the cloud size, which is sensitive to changes in compressibility as $U/t$ is varied. The cloud size evolves only for sweep times $\tau$ exceeding several milliseconds, confirming that at $\tau = \SI{1}{ms}$, the global density distribution remains frozen.

Two additional checks verify that the ramp associates existing pairs without creating new ones~\cite{Hartke2023Direct}.
First, the total unlike- ($\sum_{\delta} \expval{n_{i,\uparrow} n_{i + \delta,\downarrow}}_c$) and same-spin ($\sum_{\delta} \left( \expval{n_{i,\uparrow} n_{i + \delta,\uparrow}}_c + \expval{n_{i,\downarrow} n_{i + \delta,\downarrow}}_c \right) /2$) correlations are largely preserved by the ramp (Fig.~\ref{fig:Timescale}(b)), with only a small deviation in the opposite-spin channel for $U/t \lesssim 4$.

Second, we compare the magnetization correlations $\expval{ m_i m_{i+\delta} }_c$ before and after the ramp (Fig.~\ref{fig:Timescale}(c)).
In the noninteracting limit, the negative nonlocal correlations arising from Pauli repulsion are unaffected by the rapid quench.
In the paired regime ($U/t \gtrsim 4$), where the negative nonlocal signal originates from nonlocal pair correlations, the ramp projects the pairs onto local doublons and the off-site spin correlations vanish.
For very strong attraction ($U/t \gtrsim 16$), pairs are already local and the ramp has no significant impact.

%%%%%%%%%%%%%%%%%%%%%%%%%%%%%%%%%%%%%%%%%%%%%%%
%%%%%%%%%%%%%%% Spin Behavior %%%%%%%%%%%%%%%%%
%%%%%%%%%%%%%%%%%%%%%%%%%%%%%%%%%%%%%%%%%%%%%%%
%\section{Pairing Behavior}
\begin{figure}[hbtp]
	\centering
	\includegraphics[width=\columnwidth]{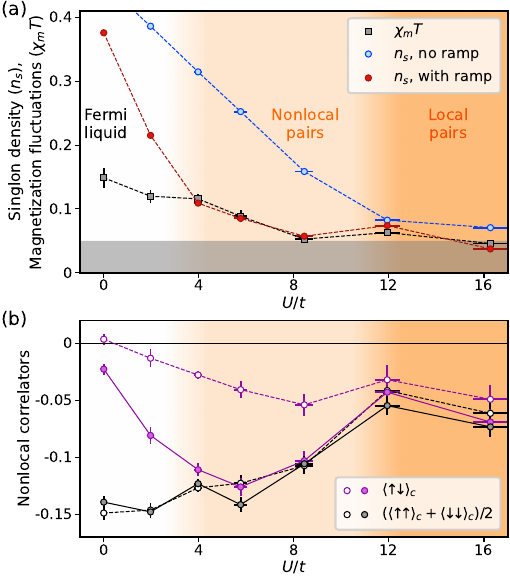}
	% \internallinenumbers
	\caption{\textbf{Pairing revealed by the rapid ramp.}
        \textbf{(a)}~Total magnetization fluctuations without the ramp ($\sum_{\delta} \expval{m_{i} m_{i + \delta}}_c$, black) together with the singlon density with (red) and without (blue) the ramp as a function of $U/t$.
        \textbf{(b)}~Nonlocal unlike-spin ($\sum_{\delta \neq 0} \expval{n_{i,\uparrow} n_{i + \delta,\downarrow}}_c$, purple) and same-spin ($\sum_{\delta \neq 0} \left(\expval{n_{i,\uparrow} n_{i + \delta,\uparrow}}_c + \expval{n_{i,\downarrow} n_{i + \delta,\downarrow}}_c \right)/2$, black) correlations with (solid markers) and without (open markers) the rapid ramp.
        }
	\label{fig:Pairing}
\end{figure}

Having established that the rapid ramp can faithfully contract nonlocal pairs into doublons, we may now use it to explore the various regimes of pairing in the attractive Hubbard gas. This is particularly fruitful in the so-called pseudo-gap phase where pairs form already above the critical temperature of superfluidity~\cite{kyung2001pairing, vsimkovic2024origin}.
These pairs are generally nonlocal in the crossover regime ($4\lesssim U/t \lesssim 8$), leading to a large singlon fraction that obscures pairing (see Fig.~\ref{fig:Pairing}(a)). Full pairing had previously been demonstrated via the absence of total magnetization fluctuations ($\sum_{\delta} \expval{m_{i} m_{i + \delta}}_c = k_B T \chi_m$, black squares in Fig.~\ref{fig:Pairing}(a)), which implies a vanishing spin susceptibility and thus a pairing gap~\cite{Hartke2023Direct}. However, this required the statistical analysis of on the order of hundreds of repetitions of the experiment. The rapid ramp now converts these nonlocal pairs into local doublons, so that a single image reveals whether the gas is fully paired (red circles in Fig.~\ref{fig:Pairing}(a)).
The remaining singlon density after the ramp can be viewed as representing the true number of unpaired fermions present before the ramp. Comparison with the magnetization fluctuations then also reveals the Fermi liquid regime at $U/t \lesssim 4$, where Pauli blocking between like spins suppresses magnetization fluctuations below that of an uncorrelated gas.

To further investigate the crossover from the Fermi liquid to the pseudogap regime of preformed pairs, we analyze the nonlocal spin correlations (Fig.~\ref{fig:Pairing}(b)).
For tightly bound pairs ($U/t \gtrsim 12$), nonlocal correlations arise from the spatial arrangement of doublons, yielding spin-independent correlations.
In the opposite extreme, for a noninteracting Fermi liquid, unlike-spin ($\uparrow \downarrow$, purple markers in Fig.~\ref{fig:Pairing}(b)) correlations vanish, whereas same-spin ($\uparrow \uparrow$ and $\downarrow \downarrow$, black markers in Fig.~\ref{fig:Pairing}(b)) correlations reflect Pauli repulsion.
In the crossover regime ($4 \lesssim U/t \lesssim 8$), characterized by nonlocal overlapping pairs, correlations arise from two competing effects: attraction between pair constituents and repulsion between distinct pairs.
This competition causes $\uparrow \downarrow$ correlations to differ from $\uparrow \uparrow$ correlations: $\sum_{\delta \neq 0} \expval{n_{i,\uparrow} n_{i + \delta,\downarrow}}_c \neq \sum_{\delta \neq 0} \expval{n_{i,\uparrow} n_{i + \delta,\uparrow}}_c$.
The rapid ramp now allows projecting these pairs onto local doublons, thereby eliminating the intra-pair contribution and leaving spin-independent correlations that purely capture the pair-pair repulsion.
Figure~\ref{fig:Pairing}(b) shows that the interaction quench reveals a transition at $U/t \sim 4$ from spin-dependent Fermi liquid behavior to the spin-independent non-local correlations expected for the paired pseudogap regime.

%%%%%%%%%%%%%%%%%%%%%%%%%%%%%%%%%%%%%%%%%
%%%%%%%%%%%%%%% Outlook %%%%%%%%%%%%%%%%%
%%%%%%%%%%%%%%%%%%%%%%%%%%%%%%%%%%%%%%%%%
%\section{Outlook}
In summary, we have established an interaction ramp technique for the attractive Hubbard gas that merges the constituents of nonlocal pairs into local doublons. This significantly enhances the visibility of charge-density-wave correlations particularly in the regime of strong correlations, when the pair size exceeds the interparticle spacing. The rapid ramp now opens new avenues for studying pairing, e.g. in spin-imbalanced systems, where pair formation and spin and charge order can be obscured not only by pair delocalization but also by spin fluctuations of the majority component. We also envision the technique to serve as a local thermometer of the gas.
Unlike established methods that rely on data-intensive nonlocal spin correlations and theoretical comparisons~\cite{xu2025neutral}, the singlon fraction after the interaction ramp is directly sensitive to the entropy and thus the temperature.
Together, simplified pair detection and local thermometry may facilitate the observation of exotic forms of pairing and superfluidity, such as the elusive Fulde--Ferrell--Larkin--Ovchinnikov (FFLO) state~\cite{feng2025search}. Analogously, interaction ramps in the dual case of the doped repulsive Hubbard model may facilitate observation of stripes or other exotic order~\cite{Xu2024Stripes}.

% Acknowledgements
We thank Tingran Wang and Zihan Xu for discussions and experimental assistance.
This work was supported by the NSF through the Center for Ultracold Atoms,
PHY-2012110 and PHY-2513210, AFOSR (FA9550-23-1-0402), DOE (DE-SC0024622), and the Vannevar Bush Faculty Fellowship (ONR N00014-19-1-2631).
C.T. acknowledges the support from the National Science Foundation Graduate Research Fellowship Program (NSF GRFP) under Grant No. 2141064.
J.H. gratefully acknowledges support through the MIT Pappalardo Fellowship.

% {\bf Author contributions:}

% {\bf Author information:}
% The authors declare no competing financial interests.

\bibliography{main}

\clearpage

%========================================
% Supplemental materials
%========================================
%TC:ignore
\setcounter{equation}{0}
\setcounter{figure}{0}
\setcounter{secnumdepth}{2}
\renewcommand{\theequation}{S\arabic{equation}}
\renewcommand{\thefigure}{S\arabic{figure}}
\renewcommand{\tocname}{Supplementary Materials}
\renewcommand{\appendixname}{Supplement}

% \tableofcontents
% \appendix

\onecolumngrid

\section*{Supplementary Information}

\subsection{Experimental setup}
\begin{figure}[hbtp]
	\centering
	\includegraphics[width=0.6\columnwidth]{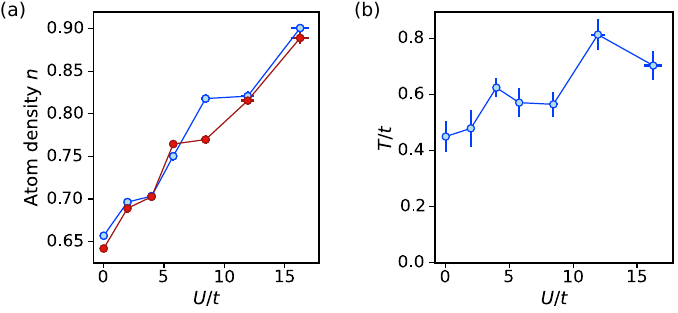}
	% \internallinenumbers
	\caption{
        \textbf{Measured density and estimated temperature of the gas.}
        \textbf{(a)}~The measured total density ($n = n_\uparrow+n_\downarrow$) of the central region of radius 12 sites of the gas both with (red, filled circles) and without (blue, open circles) performing a rapid ramp.
        \textbf{(b)}~The estimated temperature of the gas from the measured total density fluctuations and compressibility---both without the rapid ramp---using the fluctuation-dissipation theorem.~\cite{ho2009quantum, Zhou2011Universal, Hartke2020Doublon}.
	}
	\label{figSI:Density_Temperature}
\end{figure}

The attractive Fermi-Hubbard model is realized from a  degenerate gas comprised of the two lowest hyperfine states of \textsuperscript{40}K:  $\ket{F = 9/2, m_F = -9/2}$ and $\ket{F = 9/2, m_F = -7/2}$. 
The atoms occupy a single two-dimensional plane of a three-dimensional optical lattice, with in-plane spacings~$a_x \approx a_y \approx \SI{541}{nm}$ and out-of-plane spacing $a_z  \approx 3 {\;}\mu$m, as described in our previous works~\cite{Hartke2023Direct, cheuk2015quantum, Hartke2020Doublon, hartke2022quantum}.
The in-plane lattice potential is measured to be $4.3(2) E_R$ where $E_R = \hbar^2 \pi^2/(2 m a_{x,y}^2) = h \times \SI{4260}{Hz}$ is the in-plane recoil energy giving a tunneling energy $t= h \times \SI{340 \pm 20}{Hz}$.
The spatial confinement of the atoms is provided by circular box potential~\cite{nichols2019spin} and the envelope of the lattice beams with the latter providing a nearly circularly symmetric potential $(1/2) m \omega_{x,y}^2 a_{x,y}^2 r^2$ where $r$ is the radius in lattice sites and $\omega_{x,y} = 2 \pi \times \SI{26.0 \pm 0.9}{Hz}$.
The reported density and correlation data was obtained from the central region of the cloud of a radius 12 sites with nearly uniform density for each $U/t$ value (Fig.~\ref{figSI:Density_Temperature}(a)).
The temperature of the cloud is estimated from the measured total density fluctuations of the gas without the ramp in the central region and the measured compressibility extracted from the measured radial density profile of the gas and the known potential as described in Ref.~\cite{Hartke2020Doublon} (Fig.~\ref{figSI:Density_Temperature}(b)).

\subsection{Loss correction and correlators}
During the bilayer fluorescence imaging of the atoms~\cite{Hartke2023Direct, Hartke2020Doublon}, the scattered photons can eject atoms from the pinning lattice causing the measured atom distribution to be different from the original one.
For each experiment we take multiple images of the same cloud to measure the loss rate in each layer both when it is being imaged and when it is not.
The measured typical loss rate is $\SI{5 \pm 2}{\%}$ for each layer both when being imaged and when not.
Additionally, a magnetic field ramp to 195 G, below the Feshbach resonance of the two spin states (see also the Supplementary Materials of Ref.~\cite{Hartke2023Direct}), causes a 13(3) \% loss of both atoms from a doubly occupied site.
The reported density and correlation values correct for these detection losses.

The reported connected correlators of two observables $A$ and $B$ denote the joint probability of occurrence minus the product of the individual probabilities $\expval{AB}_c = \expval{AB} - \expval{A}\expval{B}$.
The reported total (nonlocal) correlations sum the correlations for distances $\abs{\delta} \leq \sqrt{10}$ including (excluding) the value at $\delta = \vec{0}$.
The precise value of the cutoff of $\delta$ has minimal effect on the total values because the correlations are vanishing above $\abs{\delta} \gtrsim 2.5$ sites.
The reported correlation maps are averaged over equivalent displacements.

\end{document}